\documentclass[iop]{emulateapj}

\newcommand*{\ac}{$A$(C)}
\newcommand*{\teff}{$T_{\rm eff}$}
\newcommand*{\logg}{$\log~g$}
\newcommand*{\feh}{[Fe/H]}

\newcommand*{\cfe}{[C/Fe]}
\newcommand*{\kms}{km s$^{-1}$}

\newcommand*{\msun}{$M_\odot$}

\newcommand*{\z}{$|Z|$}

\newcommand*{\cemps}{CEMP-$s$}

\usepackage{color}
\usepackage{epsfig}
\usepackage{graphics}
\usepackage{natbib}
\usepackage{hyperref}
\hypersetup{
    bookmarks=true,         
    unicode=false,          
    pdftoolbar=true,        
    pdfmenubar=true,        
    pdffitwindow=true,     
    pdfstartview={FitH},    
    pdftitle={},    
    pdfauthor={Lee},     
    pdfsubject={Astronomy},   
    pdfcreator={dvipdf},   
    pdfproducer={dvipdf}, 
    pdfkeywords={metal-poor stars},
    pdfnewwindow=true,      
    colorlinks=true,       
    linkcolor=red,          
    citecolor=blue,        
    filecolor=magenta,      
    urlcolor=cyan,           
    breaklinks=true,
    linktocpage
}

\shorttitle{A Carbonicity Map of the Galactic Halo}
\shortauthors{Lee et al.}
\slugcomment{Draft version, January 11, 2017}

\begin{document}

\title{Chemical Cartography. I. A Carbonicity Map of the Galactic Halo}

\author{Young Sun Lee\altaffilmark{1}, Timothy C. Beers\altaffilmark{2}, Young Kwang Kim\altaffilmark{1},
        Vinicius Placco\altaffilmark{2}, Jinmi Yoon\altaffilmark{2}, Daniela Carollo\altaffilmark{3,4},
        Thomas Masseron\altaffilmark{5}, and Jaehun Jung\altaffilmark{6}}
\altaffiltext{1}{Department of Astronomy and Space Science, Chungnam National University,
                 Daejeon 34134, Korea; youngsun@cnu.ac.kr}
\altaffiltext{2}{Department of Physics and JINA Center for the Evolution of the Elements, University
	         of Notre Dame, Notre Dame, IN 46556, USA}
\altaffiltext{3}{Research School of Astronomy and Astrophysics, The Australian National University,
                 Canberra, ACT 2611, Australia}
\altaffiltext{4}{INAF--Osservatorio Astronomico di Torino, Strada Osservatorio 20, Pino Torinese, I-10020, Italy}
\altaffiltext{5}{Institute of Astronomy, University of Cambridge, Madingley Road, Cambridge CB3 0HA, UK}
\altaffiltext{6}{Department of Astronomy, Space Science, and Geology, Chungnam National University,
                 Daejeon 34134, Korea}

\begin{abstract}

We present the first map of carbonicity, [C/Fe], for the halo system of
the Milky Way, based on a sample of over 100,000 main-sequence turnoff
stars with available spectroscopy from the Sloan Digital Sky Survey.
This map, which explores distances up to 15 kpc from the Sun, reveals
clear evidence for the dual nature of the Galactic halo, based on the
spatial distribution of stellar carbonicity. The metallicity
distribution functions of stars in the inner- and outer-halo regions of
the carbonicity map reproduce those previously argued to arise from
contributions of the inner- and outer-halo populations, with peaks at
\feh\ = $-1.5$ and $-2.2$, respectively. From consideration of the
absolute carbon abundances for our sample, $A$(C), we also confirm that
the carbon-enhanced metal-poor (CEMP) stars in the outer-halo region
exhibit a higher frequency of CEMP-no stars (those with no
overabundances of heavy neutron-capture elements) than of CEMP-$s$ 
stars (those with strong overabundances of elements associated
with the $s$-process), whereas the stars in the inner-halo region
exhibit a higher frequency of CEMP-$s$ stars. We argue that the contrast
in the behavior of the CEMP-no and CEMP-$s$ fractions in these regions
arises from differences in the mass distributions of the mini-halos from
which the stars of the inner- and outer-halo populations formed, which
gives rise in turn to the observed dichotomy of the Galactic halo.

\end{abstract}


\section{Introduction}

Observations of the kinematics and chemistry of stars in the halo of the
Milky Way (MW) provide valuable clues to its assembly history, and that
of other large galaxies in general. Until relatively recently, the MW's
diffuse stellar halo has been thought to comprise only a single stellar
population -- stars with similar ages, chemical abundances, and
kinematics. This long-standing idea has been challenged in the past
decade by both observations (e.g., \citealt{carollo2007, carollo2010,
dejong2010, beers2012, an2013, an2015, hattori2013, allendeprieto2014,
chen2014, fernandez2015, fernandez2016a, fernandez2017, janesh2016, das2016}) and
ever more sophisticated simulations of the formation of large MW-like
galaxies (e.g., \citealt{zolotov2009, font2011, mccarthy2012,
tissera2013, tissera2014}). We briefly summarize our current
understanding below.

\subsection{The Nature of the Galactic Halo}

The evidence presented to date indicates that the diffuse Galactic halo
comprises at least two distinct stellar components -- the inner-halo
population (IHP) and the outer-halo population (OHP), which can be
distinguished based on their different spatial distributions,
metallicity ([Fe/H]\footnote[7]{We use the terms metallicity and [Fe/H]
interchangeably throughout this paper.}), and kinematics. As summarized
by \citet{carollo2010} and \citet{beers2012}, the inner-halo
component dominates the population of halo stars found at distances up
to 10--15 kpc from the Galactic center, while the outer-halo component
dominates in the region beyond 15--20 kpc. In addition, the inner halo
exhibits a flatter density profile than the nearly spherical outer halo.
The metallicity distribution function (MDF) of inner-halo
stars peaks at [Fe/H] = --1.6, while the outer-halo stars exhibit a peak
at [Fe/H] = --2.2. Kinematically, the IHP shows either zero or slightly
prograde rotation with respect to the Galactic center, with stars on
somewhat eccentric orbits, while the OHP exhibits a net retrograde
rotation of about --80 \kms, with stars on more circular orbits
(\citealt{carollo2007, carollo2010, kinman2012, hattori2013}). The observed 
velocity ellipsoids of the populations also differ, in the sense that
the OHP is kinematically ``hotter'' than the IHP \citep{carollo2007, carollo2010, 
carollo2014, an2015, helmi2017}.

This view has been challenged by claims that systematic errors in
adopted distances may artificially induce an apparent dual nature of the
halo for samples whose kinematics are deduced from relatively local
stars (\citealt{schonrich2011}). However, \citet{beers2012} refuted this
argument, based on a number of lines of evidence. Furthermore,
\citet{das2016}, using a completely different modeling approach for
distant K-type giants from the Sloan Digital Sky Survey (SDSS;
\citealt{york2000}), come to essentially the same conclusions as those
arising from the interpretation of the local kinematics of halo stars.
Most recently, \citet{helmi2017} have used parallaxes and proper motions
from the first Gaia data release \citep{gaia2016}, in combination with
metallicity estimates from Data Release 5 of the RAVE survey \citep{kunder2016}, to
provide confirming evidence that the retrograde signature for OHP stars
is highly unlikely to be an artifact of incorrect distance estimates in
previous work.

Claims that the selection function of metal-poor stars in local samples
may have led to an incorrect dual-halo inference (echoed by, e.g.,
\citealt{monachesi2016}) have been empirically resolved by the results
of \citet{an2013, an2015}. These authors estimated metallicities from
coadded SDSS $ugriz$ photometry for stars with distances in the range
5--10 kpc from the Sun, and derived an MDF with two peaks (one at [Fe/H]
$\sim -1.4$ and the other at [Fe/H] $\sim -1.9$). They associated the
more metal-rich component with the IHP and the more metal-poor component
with the OHP. They further demonstrated that the two populations exhibit
the expected kinematic signatures inferred from the analysis of
\citet{carollo2007, carollo2010}. Since no spectroscopic selection is
made in the analysis by An et al., the argument that the dual-halo
interpretation could arise from such a source is rendered moot.

\subsection{Carbon-enhanced Metal-poor Stars as Probes of the Galactic
Halo System}

Among the stars of the halo system, the so-called carbon-enhanced
metal-poor (CEMP; \citealt{beers2005}) stars have emerged as important
tracers of the assembly and star-formation history of the Galactic halo,
as summarized briefly here.

It has been recognized for over a decade that CEMP stars can be divided
into several subclasses, according to their level of neutron-capture
element enhancement -- CEMP-no, CEMP-$s$, CEMP-$r$, and CEMP-$r/s$
\citep{beers2005}. CEMP-no stars exhibit no overabundances of heavy
neutron-capture elements, while CEMP-$s$ objects show strong
overabundances of elements associated with the $s$-process, such as Ba.
CEMP-$r$ stars possess strong enhancements of $r$-process elements
such as Eu.  Carbon-enhanced stars with possible contributions from
both the $r$-process and the $s$-process have been classified
as CEMP-$r/s$ stars\footnote[8]{Recently, \citet{hampel2016} have
presented compelling evidence that the CEMP-$r/s$ stars are more likely
associated with a proposed intermediate neutron-capture process, the
$i$-process.}.

The ever-growing numbers of CEMP stars with available high-resolution
spectroscopic results have shown that the CEMP-$s$ and CEMP-no stars are
the most populous subclasses, accounting for more than 95\% of the
samples of carbon-enhanced stars at low metallicity. Numerous studies
have also shown that the frequency of CEMP stars in the halo system
dramatically increases with decreasing metallicity (e.g.,
\citealt{lucatello2006, lee2013, yong2013, placco2014}), as well as with
distance from the Galactic plane (\citealt{frebel2006, carollo2012,
beers2016}). It has also become clear that most CEMP-no stars have [Fe/H]
$<$ --2.5, while CEMP-$s$ stars are predominantly found among
with [Fe/H] $>$ --2.5 (e.g., \citealt{aoki2007,
yoon2016}). Long-term radial-velocity monitoring of CEMP stars (e.g.,
\citealt{starkenburg2014, hansen2016a, hansen2016b, jorissen2016}) has provided
strong evidence for different binary fractions associated with the
CEMP-$s$ and CEMP-no stars -- $\sim$82\% of CEMP-$s$ (including
CEMP-$r/s$) stars are binaries, while \textbf{only} $\sim$17\% of CEMP-no stars are
binaries, consistent with the observed binary fraction of other
metal-poor stars in the halo \citep{carney2003}. It is thus clear
that binarism is not required to account for the origin of the CEMP-no
stars; their distinctive chemical signatures are very likely the result
of the pollution of their natal clouds by first-generation stars.

From a kinematic analysis of CEMP stars in SDSS, \citet{carollo2012}
reported a higher fraction of CEMP stars associated with the OHP than 
with the IHP. From a similar analysis of the limited number of
stars with available high-resolution spectroscopic classifications of
CEMP stars, \citet{carollo2014} demonstrated that the OHP exhibits a
relatively higher fraction of CEMP-no stars than found for CEMP stars
associated with the IHP, which slightly favored CEMP-$s$ stars over
CEMP-no stars.

The goal of this study is twofold. We first explore further evidence
for the dichotomy of the diffuse halo system, based on the observed {\it
in situ} spatial distribution of a large sample of $\sim$~105,700
main-sequence turnoff (MSTO) stars from SDSS with measured
carbon-to-iron ratios ([C/Fe], which we refer to as ``carbonicity''),
located within 15 kpc of the Sun. Secondly, we search for the origin of
the duality of the Galactic halo by assessing the relative fractions of
CEMP-no and CEMP-$s$ stars in the chemically separated IHP and OHP,
making use of a recently suggested scheme based on absolute carbon
abundances, $A$(C)\footnote[9]{The conventional notation is used,
\ac = log\,$\epsilon$(C) = log\,($N_{\rm C}$/$N_{\rm H}) + 12$.}, developed
by \citet{yoon2016}, to distinguish CEMP-no stars from CEMP-$s$ stars
based on medium-resolution spectroscopy alone. Note, however, 
that our MSTO sample covers substantially different ranges
in metallicity and luminosity than the sample of \citet{yoon2016}. 
Yoon et al.'s sample is dominated by metal-poor giants with 
[Fe/H] $<$ --2.0, while our MSTO stars primarily cover [Fe/H] $>$ --2.5. 
As is addressed in Section 4.4, these different samples suggest 
we must apply different reference 
points of the critical \ac\ value to separate CEMP-no from CEMP-$s$ stars than 
that (\ac\ = 7.1) of \citet{yoon2016}.

This paper is arranged as follows. Section 2 summarizes the criteria
used to select our program stars. In Section 3, we validate our
measurement of [C/Fe] by comparison with available high-resolution
spectroscopy and noise-added synthetic spectra. In Section 4, we obtain
the spatial distribution of \cfe\ for the stars in our sample, and identify four
regions that we associate with different stellar populations in the MW. The
MDFs and the carbonicity distribution functions (CDFs) for each region
are considered. We also assess the fractions of the CEMP-no stars and
CEMP-$s$ stars for these regions. Section 5 presents a plausible origin
of the dichotomy of the Galactic halo, based on the data in hand.
A summary of our results, and prospects for future work, is presented in
Section 6.

\begin{figure}[t]
\centering
\plotone{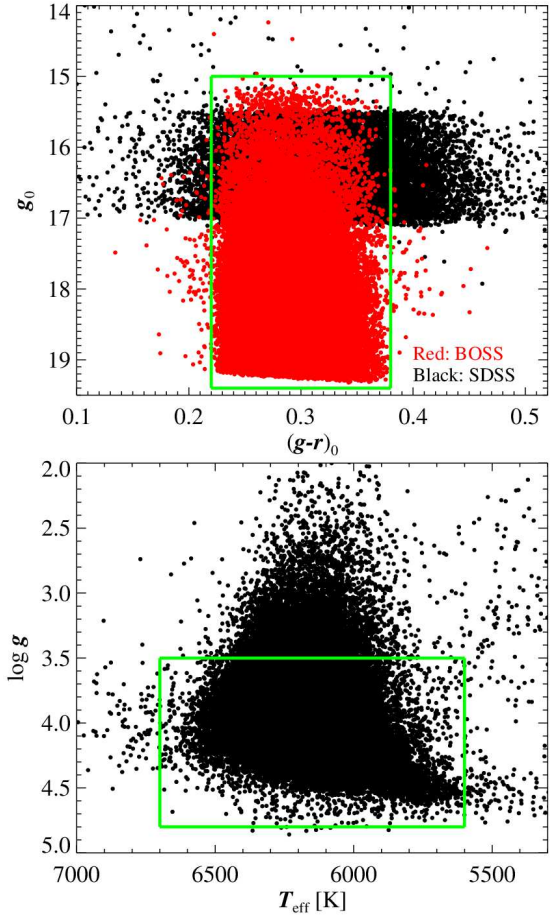}
\caption{Distribution of $g_{\rm 0}$ vs. $(g-r)_{\rm 0}$ (top panel)
and \logg\ vs. \teff\ (bottom panel) for
spectrophotometric standard stars from SDSS. The red dots represent
stars from BOSS, while the black dots are stars from legacy SDSS,
SEGUE-1, and SEGUE-2. The green boxes in the top and bottom panels
indicate the magnitude and color cuts and the surface gravity and
temperature cuts used to select our sample stars, respectively. Note
that only spectrophotometric standard stars are shown in this figure;
our final selected sample comprises all stars from the legacy SDSS,
SEGUE-1, SEGUE-2, and BOSS surveys that satisfy these criteria.}
\label{fig:sample}
\end{figure}

\section{Sample Selection}

The sample used in this study primarily comprises medium-resolution ($R
\sim$ 2000) spectra for stars observed during the course of the legacy
SDSS program and the Sloan Extension for Galactic Understanding and
Exploration (SEGUE-1; \citealt{yanny2009}), as well as SEGUE-2 (C. Rockosi
et al. 2017, in preparation). We also included spectra of spectrophotometric
standard stars obtained during the Baryon Oscillation Spectroscopic
Survey (BOSS; \citealt{dawson2013}), which was one of the four projects
executed during the third phase of the SDSS \citep{alam2015}.

We selected stellar spectra from legacy SDSS, SEGUE-1, SEGUE-2, and
BOSS, based on $g_{\rm 0}$ magnitude, $(g-r) _{\rm 0}$ color, surface
gravity (\logg), and effective temperature (\teff) that satisfied the
following conditions: 15.0 $ \leq g_{0} \leq 19.4$, 0.22 $\leq (g-r)_{0}
\leq 0.38$, 3.5 $\leq$ \logg\ $\leq$ 4.8, and 5600 K $\leq $ \teff\
$\leq 6700$~K. The magnitude and color cuts are similar to those for the
selection of spectrophotometric standard stars in BOSS, which correspond
to the MSTO objects of an old stellar population; the gravity and
temperature ranges also restrict our sample to MSTO stars. The green boxes in
Figure~\ref{fig:sample} indicate the ranges of magnitude, color,
surface gravity, and temperature used to select our program stars.

Estimates of the stellar atmospheric parameters (\teff, \logg, and \feh)
for our program stars were obtained through application of the SEGUE
Stellar Parameter Pipeline (SSPP; \citealt{lee2008a, lee2008b,
allendeprieto2008, lee2011, smolinski2011}). The carbonicity, \cfe, was
estimated following the procedures of \citet{lee2013}. Typical errors in
the stellar parameters are $\sim 180$~K for \teff, $\sim 0.24$ dex for
\logg, and $\sim 0.23$ dex for \feh, respectively, while the error in
the estimated [C/Fe] is better than 0.35 dex in the temperature range
4400~K $\leq$ \teff\ $\leq$ 6700~K for spectra with signal-to-noise
ratios S/N $\geq 15$ \AA$^{-1}$ \citep{lee2013}, increasing to $\sim
0.5$ dex at S/N = 10 \AA$^{-1}$. More recent tests of errors in the SSPP
estimates of carbonicity suggest that these estimates are conservative,
and that the actual errors in their determination are likely to be
$\sim$0.25--0.35 dex over these ranges in S/N.

We removed stars from our sample that were drawn from spectroscopic
plug-plates taken in the direction of open cluster and globular cluster
fields, in order to minimize contamination. Since there are many stars
that were observed more than once, we chose to include only 
the spectrum with the highest S/N for multiply observed
stars. We visually inspected all the spectra of stars with [Fe/H]
$\le$ --1.0 used in our program to exclude cool white dwarfs and objects
with defects in their spectra, which could lead to spurious estimates of 
atmospheric parameters. For a clear detection of the CH $G$-band around
4300\,{\AA}, which ensures a good measurement of [C/Fe], we selected
spectra for stars with equivalent widths of the carbon feature in the
region 4290--4318\,{\AA} larger than 0.6\,{\AA}, because we have found that
spurious estimates of [C/Fe] become significant below this value.

We followed the methodology of \citet{beers2000, beers2012} to determine
the distance to each star; the typical distance uncertainty is of the
order of 15-20\%. This was then used to compute the distance from the
Galactic mid-plane, $Z$, and the distance from the Galactic center
projected onto the plane, $R$, assuming the Sun is located at 8.5 kpc
from the Galactic center.

Our final sample comprises stars with spectra having S/N $\ge$ 12
\AA$^{-1}$ (see the next section for justification of this cut) and
valid estimates of \teff, \logg, \feh, \cfe, $Z$, and $R$. The total
number of stars in the sample is $N \sim 105,700$.

\begin{figure}[t]
\centering
\plotone{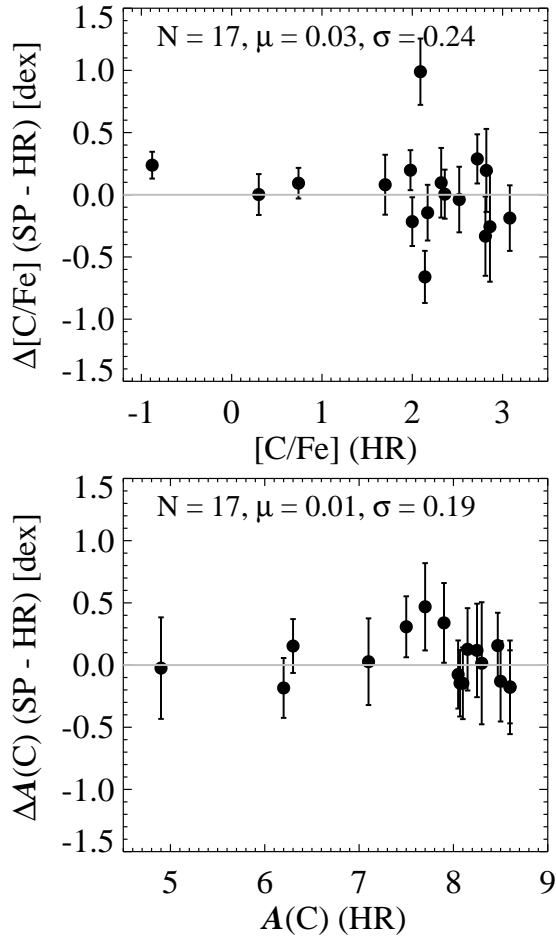}
\caption{Distribution of residuals for SSPP-estimated [C/Fe] (top panel)
and \ac\ (bottom panel) as a function of the high-resolution
determinations. `SP' indicates the estimates from the SSPP; `HR' refers
to the literature values from \citet{yoon2016}. After correction for a
$+$0.16 dex offset in the SSPP estimates (already performed for the data
shown in the figure; see text), there are no significant trends in the
residuals. The legend for each panel lists the total number of
comparison stars ($N$), the mean offset ($\mu$), and the standard
deviation ($\sigma$). The error bars are calculated by adding the
uncertainties of the SSPP and HR estimates in quadrature. We assumed
that the typical uncertainty of the HR [C/Fe] determination is 0.1,
and 0.2 dex for \ac. The uncertainty for the SSPP estimate is that
provided by the pipeline.}
\label{fig:ccheck}
\end{figure}

\begin{figure}[t]
\centering
\plotone{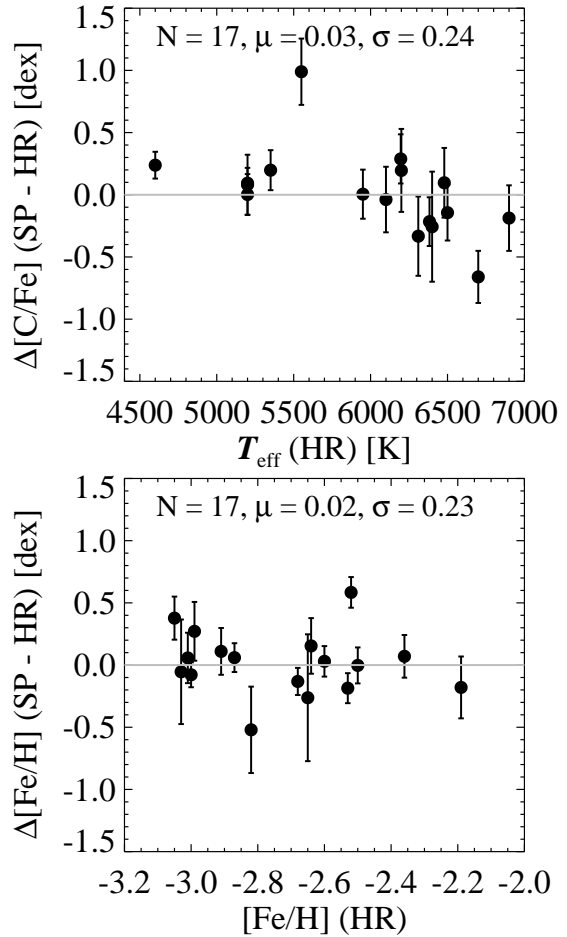}
\caption{Top panel: Distribution of residuals for SSPP-estimated
[C/Fe], as a function of \teff, after correction for the $+$0.16 dex
offset in the SSPP estimates. `SP' and `HR' are as defined in Figure~\ref{fig:ccheck}.
No significant trend of the SSPP [C/Fe]
residuals with \teff\ is found. Bottom panel: Distribution of residuals
for SSPP-estimated [Fe/H], as a function of the HR metallicity estimates; no corrections
were needed with respect to the HR determinations. No trend with [Fe/H]
is seen. The error bars are calculated by adding the uncertainties of the SSPP and HR
estimates in quadrature. We assumed that the typical uncertainty of both
HR [C/Fe] and HR [Fe/H] determinations is 0.1 dex. The uncertainty for
the SSPP estimate is that provided by the pipeline.}
\label{fig:tcheck}
\end{figure}

\begin{figure*}[t]
\centering
\plotone{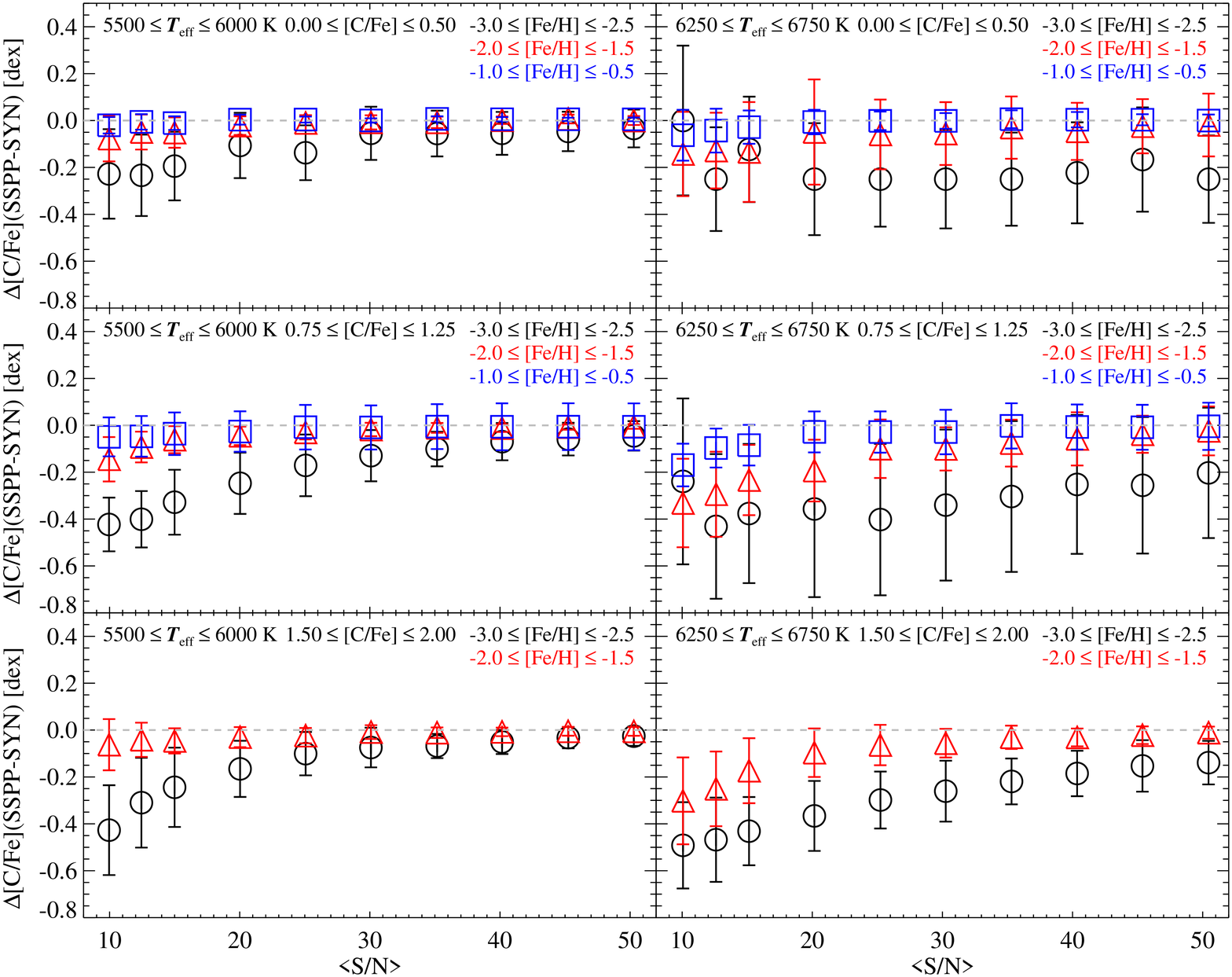}
\caption{Distribution of residuals in [C/Fe] between the SSPP (denoted
as `SSPP') estimates and values obtained for noise-added synthetic spectra
(indicated by `SYN'). $\langle$S/N$\rangle$ is an average signal-to-noise
ratio per angstrom. The ranges of \teff, \feh, and \cfe\ are listed in the
legend of each panel. Each symbol is represented by the same color as the
metallicity range. The error bar is represented by the standard deviation
for the stars in each bin of S/N.}
\label{fig:snr}
\end{figure*}

\section{Check on Estimate of [C/Fe] for Warm,  Metal-poor Stars}

\subsection{Comparison with High-resolution Spectroscopy}

The BOSS calibration stars in our sample are extremely useful probes of
the Galactic halo, because they reach two magnitudes fainter than the
legacy SDSS and SEGUE calibration stars (see Figure~\ref{fig:sample}).
Although inclusion of redder stars would add a large
number of more distant giants, one clear advantage of our MSTO sample is
that stars in this stage of evolution do not suffer from dilution of
their surface carbon abundances by material that has undergone CN
cycling (due to extra mixing occurring along the red-giant branch).

We note, however, that this choice also introduces some difficulty for
the detection of molecular carbon features for a subset of the MSTO
stars in our sample, in particular for warmer and lower-metallicity
stars. This follows because the strength of the CH $G$-band around
4300\,{\AA}, which is used by the SSPP to estimate [C/Fe], decreases at
higher temperatures and lower metallicity. As a result, some bona-fide
CEMP stars may escape detection, and will be misclassified
as carbon-normal stars. This problem occurs even for high-resolution
spectroscopic analyses -- examples of such cases are the well-known
warm, metal-poor subdwarfs G~64-12 ([Fe/H] = --3.29) and G~64-37 ([Fe/H]
= --3.11), whose status as CEMP-no stars was only recently demonstrated
by \citet{placco2016}, based on extremely high S/N ($\sim700$),
high-resolution ($R\sim95,000$) spectra.

In order to check the reliability of the SSPP estimate of carbonicity,
we first compared our [C/Fe] values with the results from high-resolution
spectroscopy for the small sample of stars in common from Table 1 of
\citet{yoon2016}. Figure~\ref{fig:ccheck} shows the results of this
exercise. The top panel of this figure is a residual plot of the
SSPP-estimated [C/Fe] compared to the literature values, where
`SP' indicates the estimation from the SSPP, and `HR'
indicates the literature value from \citet{yoon2016}. We identified a positive
offset between the SSPP and literature estimates of [C/Fe] of $+$0.16 dex, and have
reduced the plotted SSPP carbonicity estimates by this amount. The bottom panel
is a similar residual plot, but for \ac\footnote[10]{$A$(C) = log\,
$\epsilon$(C) is obtained from the medium-resolution determinations
using $A$(C) = [C/Fe] $+$ [Fe/H] $+$ $A$(C) $_{\odot}$, where we adopt the
solar abundance of carbon from \citet{asplund2009}, $A$(C)$_{\odot}$ =
8.43.}. The offset of $+$0.16 dex in [C/Fe] is also corrected in this
panel.

The top panel of Figure \ref{fig:tcheck} shows the distribution of the
carbonicity residuals as a function of \teff\ (corrected for the $+$0.16 dex
offset as before). The bottom panel is the distribution of metallicity residuals as
a function of the HR literature estimates of [Fe/H];  we found no significant
metallicity offset of the SSPP estimates from the high-resolution
values. In summary, Figures~\ref{fig:ccheck} and \ref{fig:tcheck} do not exhibit
any significant trends of the SSPP [C/Fe] and [Fe/H] estimates against
the high-resolution estimates, and show good agreement with very small
scatter, even in the range of [Fe/H] $<$ --2.5. Most of the comparison
stars fall in the same warm, low-metallicity regime as for our sample
stars; it appears that we are able to obtain reliable estimates of
[C/Fe] using the SSPP for these stars.

\subsection{Impact of Signal-to-noise Ratio on [C/Fe]}

The difficulty of detecting molecular carbon features for warmer and
lower-metallicity stars in our sample becomes worse for low-S/N spectra,
and may also result in underestimates of [C/Fe], even when the carbon
features are detected. Below we describe a set of tests we have carried
out to quantify these effects.

We first injected noise into the grid of synthetic spectra that was used
to estimate [C/Fe]. We then chose an SDSS/SEGUE spectrum that has similar
values of \teff\ and \feh\ to a model spectrum to which we want to add
noise, and created a noise vector by dividing the flux of the model
spectrum by the S/N values of the selected SDSS/SEGUE spectrum as a
function of wavelength. Following this step, we injected the generated
noise into the synthetic spectra and obtained noise-added synthetic
spectra with S/N = 10.0, 12.5, 15.0, 20.0, 25.0, 30.0, 35.0, 40.0, and
50.0 \AA$^{-1}$, and simulated 25 different noise-added synthetic
spectra at each S/N value. Finally, we processed the noise-added spectra
through the SSPP to determine estimates of [C/Fe], holding the \teff\
and \logg\ values that represent the synthetic spectra fixed, and letting
only \feh\ and \cfe\ vary (as is done for the usual operation of the
SSPP on program stars) to minimize the $\chi^{2}$ values.

\begin{figure}
\centering
\plotone{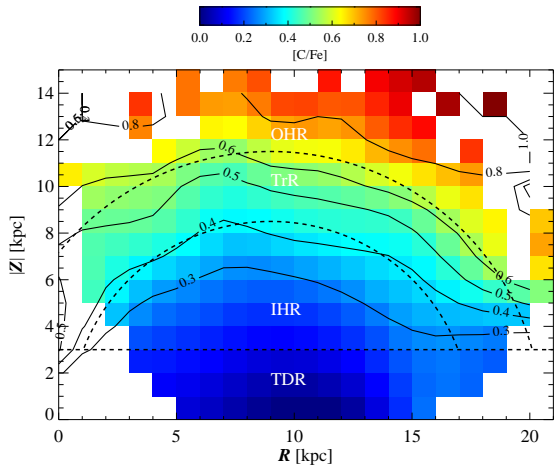}
\caption{Map of carbonicity, [C/Fe], for our MSTO sample in the plane of \z\
vs. $R$. \z\ is the absolute distance from the Galactic mid-plane,
while $R$ is the distance from the Galactic center projected onto the
plane, $R$. The bin size is 1$\times$1 kpc; each pixel contains at least
three stars. The color code is based on the median value of \cfe\ in
each pixel, with the color scale shown above the plot. Contours of
median [C/Fe] values are overplotted. The dashed line at \z\ = 3 kpc
indicates the approximate upper boundary of a thick-disk region (TDR),
while the area between the upper limit of the thick disk and the inner dashed
circle represents an inner-halo region (IHR). The area above the outer
dashed circle is assigned to an outer-halo region (OHR). The area
between the inner and outer dashed circles is associated with a
transition region (TrR), where stars from both the IHP and OHP are
expected to be found. Each region is abbreviated in white letters. We
have applied a Gaussian kernel to the map to obtain a smooth
distribution of [C/Fe]. We decided to force stars with [C/Fe] $>$ $+$1.0
into the [C/Fe] = $+$1.0 bin, and stars with [C/Fe] $<$ 0.0 into the [C/Fe] =
0.0 bin, in order to better illustrate the subtle contrast in the map.}
\label{fig:cmap}
\end{figure}

We proceeded by grouping the noise-added synthetic spectra into two
regions of \teff\ (5500--6000~K and 6250--6750~K), three regions
of \feh\ (--1.0 to --0.5, --2.0 to --1.5, and --3.0 to --2.5), and three
regions of \cfe\ ($+$0.00 to $+$0.50, $+$0.75 to $+$1.25, $+$1.50 to
$+$2.00). We only considered one gravity range, 3.5 $\leq$ \logg\ $\leq$
4.5. These ranges of temperature, gravity, and metallicity well represent
the parameter space of our MSTO sample. For each S/N value, we then obtained 
the median and standard deviation of the differences in \cfe\ estimates between the
SSPP-estimated values and the model values for each group of spectra.
We adopted a median rather than a mean value to prevent the residuals from being 
dominated by a few spurious determinations of \cfe, especially for 
low-S/N spectra. 

Figure~\ref{fig:snr} shows how the residuals in
[C/Fe] change with S/N. The ranges of temperature, metallicity, and carbonicity 
considered are listed at the top of each panel; the symbols have
the same colors as the metallicity ranges. The label `SSPP' indicates the
SSPP values, whereas `SYN' denotes the model values. In the figure,
$\langle$S/N$\rangle$ is an average S/N per angstrom. The error bar
represents the standard deviation.

Figure~\ref{fig:snr} indicates that our technique for determining [C/Fe]
reproduces it reasonably well, because the median difference between the
SSPP and the model is mostly less than 0.3 dex, commensurate with the
uncertainty of the SSPP [C/Fe] estimate, for [Fe/H] $\ge$ --2.0 and S/N
$\ge 12.5$ \AA$^{-1}$. Larger deviations exist for stars with --3.0
$\leq$ [Fe/H] $\leq$ --2.5 and/or S/N $<20$. Based on inspection of
Figure~\ref{fig:snr}, we adjusted our estimate of [C/Fe] by 0.1 or 0.2
dex for stars in the following ranges:

\begin{itemize}
\item 5600 $\leq$ \teff\ $<$ 6100 K, --2.5 $<$ [Fe/H] $\leq$ --2.0, 12.0 $\leq$ S/N $\leq$ 20.0 ($+$0.1 dex correction)
\item 5600 $\leq$ \teff\ $<$ 6100 K, [Fe/H] $\leq$ --2.5, S/N $\ge$ 12.0 ($+$0.2 dex correction)
\item 6100 $\leq$ \teff\ $\leq$ 6700 K, --2.0 $\leq$ [Fe/H] $\leq$ --1.5, 12.0 $\leq$ S/N $\leq$ 20.0 ($+$0.1 dex correction)
\item 6100 $\leq$ \teff\ $<$ 6700 K, [Fe/H] $\leq$ --2.5, S/N $\ge$ 12.0 ($+$0.2 dex correction)
\end{itemize}

\noindent The numbers of stars affected are is 468, 54, 6676, and 1461, respectively.
Stars for which the molecular carbon features are not detected fall out
of our sample and are no longer considered.

One unexpected trend, seen in the first and second panels of the
right-hand column of plots (the warmest stars), is that the residuals for the
lowest-metallicity stars suddenly decrease at S/N = 10.0, in contrast
to the behavior at higher S/N. This may be caused by a very broad and
poorly constrained distribution of $\chi^{2}$ values at low S/N, high
\teff, and low \feh, which results in an unreliable estimate of \cfe. For
this reason we decided to cut our sample by requiring S/N $\ge$ 12.0.
Using a much larger grid of noise-added synthetic spectra, we are
currently developing a procedure to derive a smooth correction function
of SSPP-derived [C/Fe] for MSTO objects that could suffer from this
problem, as functions of S/N, \teff, \feh, and \cfe\ of the observed
medium-resolution spectra.

\begin{figure}
\centering
\plotone{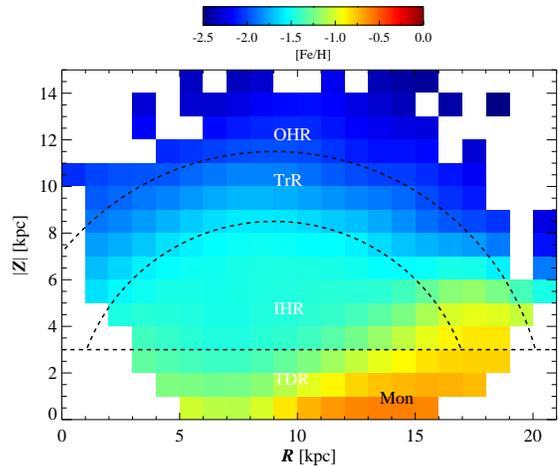}
\caption{Same as in Figure \ref{fig:cmap}, but for metallicity, [Fe/H].
It is clear that the regions (abbreviated in white letters) selected to
demarcate the levels of carbonicity in Figure~\ref{fig:cmap} also
correspond to different levels of metallicity. Note, in particular, the
contrast between the IHR, with [Fe/H] $\sim$ --1.5, and the OHR, with
[Fe/H] $<$ --2.0. We have applied a Gaussian kernel to the map to obtain
a smooth distribution of [Fe/H]. We decided to force stars with [Fe/H]
$<$ --2.5 into the [Fe/H] = --2.5 bin, and stars with [Fe/H] $>$ 0.0
into the [Fe/H] = 0.0 bin, in order to better illustrate the subtle
contrast in the map. The Monoceros stream, with [Fe/H] $\sim$ --1.0,
appears as a yellow-orange region at $R > 10$ kpc and \z\ $<$ 5 kpc, and
is labeled as ``Mon''.}
\label{fig:fmap}
\end{figure}

\section{Results}

Having prepared our MSTO sample as described above, and carried out a
number of small corrections to our estimates of [C/Fe], we now consider
the distribution of carbonicity (Sec. 4.1) and  metallicity (Sec. 4.2).
We then consider the apparent dichotomy of the halo system of the Galaxy in
metallicity (Sec. 4.3) and in absolute carbon abundance, $A$(C) (Sec.
4.4), over distances extending to 15 kpc from the Sun.

\subsection{Spatial Distribution of Carbonicity, [C/Fe]}

Figure \ref{fig:cmap} is a map of \cfe\ for our MSTO sample in the plane 
of \z\ vs. $R$. From inspection, it is clear that, at
a given $R$, the median \cfe\ (indicated by the contour lines) gradually
increases with increasing \z. The dashed circles in the figure are
centered at $R=$ 9 kpc, the inner circle having a radius of 8.5 kpc,
while the outer circle has a radius of 11.5 kpc. The inner circle falls
close to the contour of [C/Fe] = $+$0.4, whereas the outer circle
closely follows the contour with [C/Fe] = $+$0.6. Making use of these
contour lines as fiducials to guide our choice, we divide the map into
four primary regions, and associate stars in each region with an
expected dominant stellar population as follows:

\begin{itemize}

\item Thick-disk region (TDR) -- The region
with \z\ $\leq 3.0$ kpc shown in the figure. Stars in this region
are likely to be dominated by the thick-disk population (TDP).

\item Inner-halo region (IHR) -- The region corresponding to
\z\ $>$ 3 kpc and the inner dashed circle, which closely follows the [C/Fe] = $+$0.4 contour.
Stars in this region are expected to be dominated by the IHP.

\item Outer-halo region (OHR) -- The region beyond the outer dashed
circle. Stars in this region are expected to be dominated by the OHP.

\item Transition region (TrR) -- Defined as the region between the two
dashed circles. Stars in this region are expected to be contributed by
both the IHP and OHP; for convenience we refer to these stars as a
transition population (TrP), although it is understood that this
represents an overlap of the IHP and OHP, not a separate population.

\end{itemize}

\begin{figure*}
\centering
\plottwo{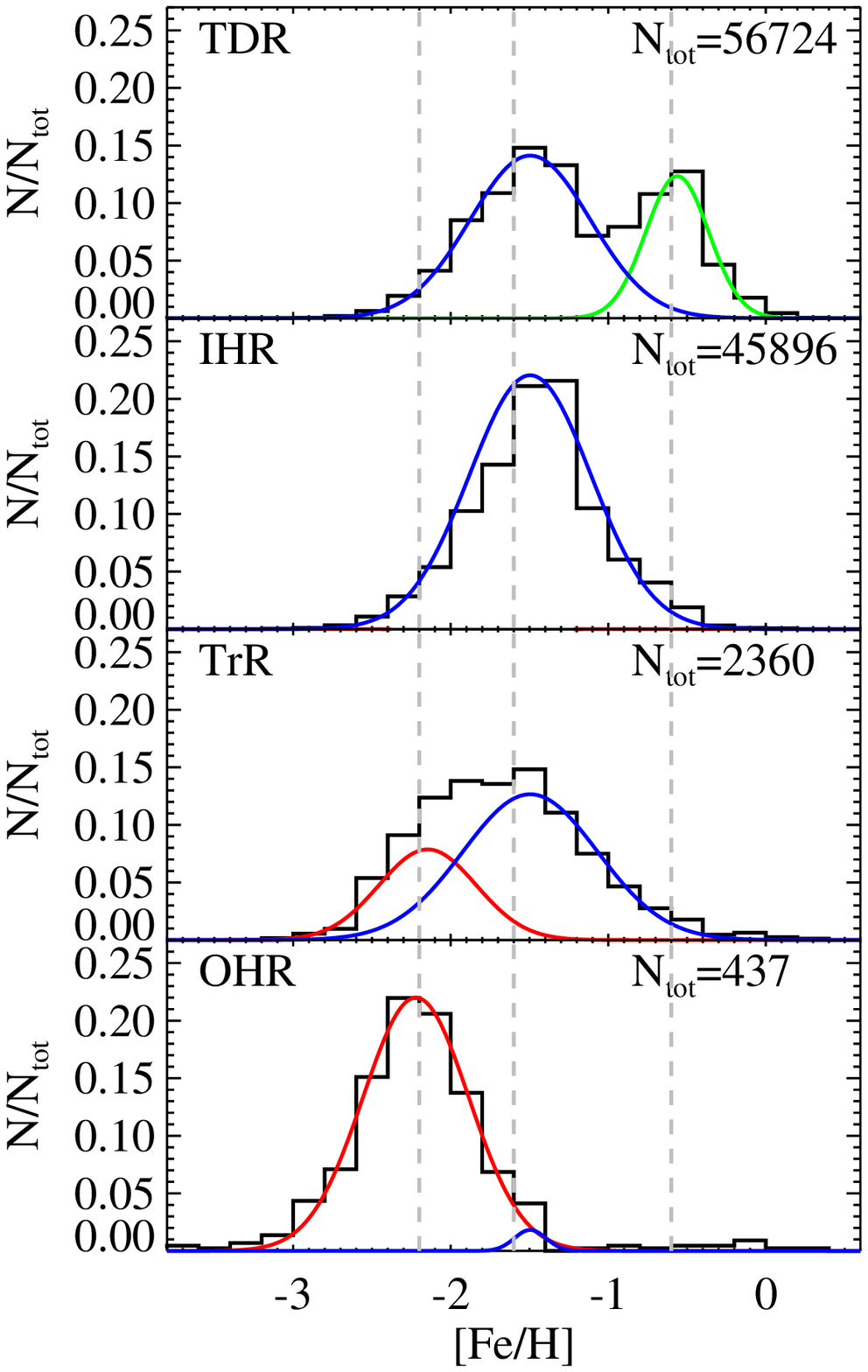}{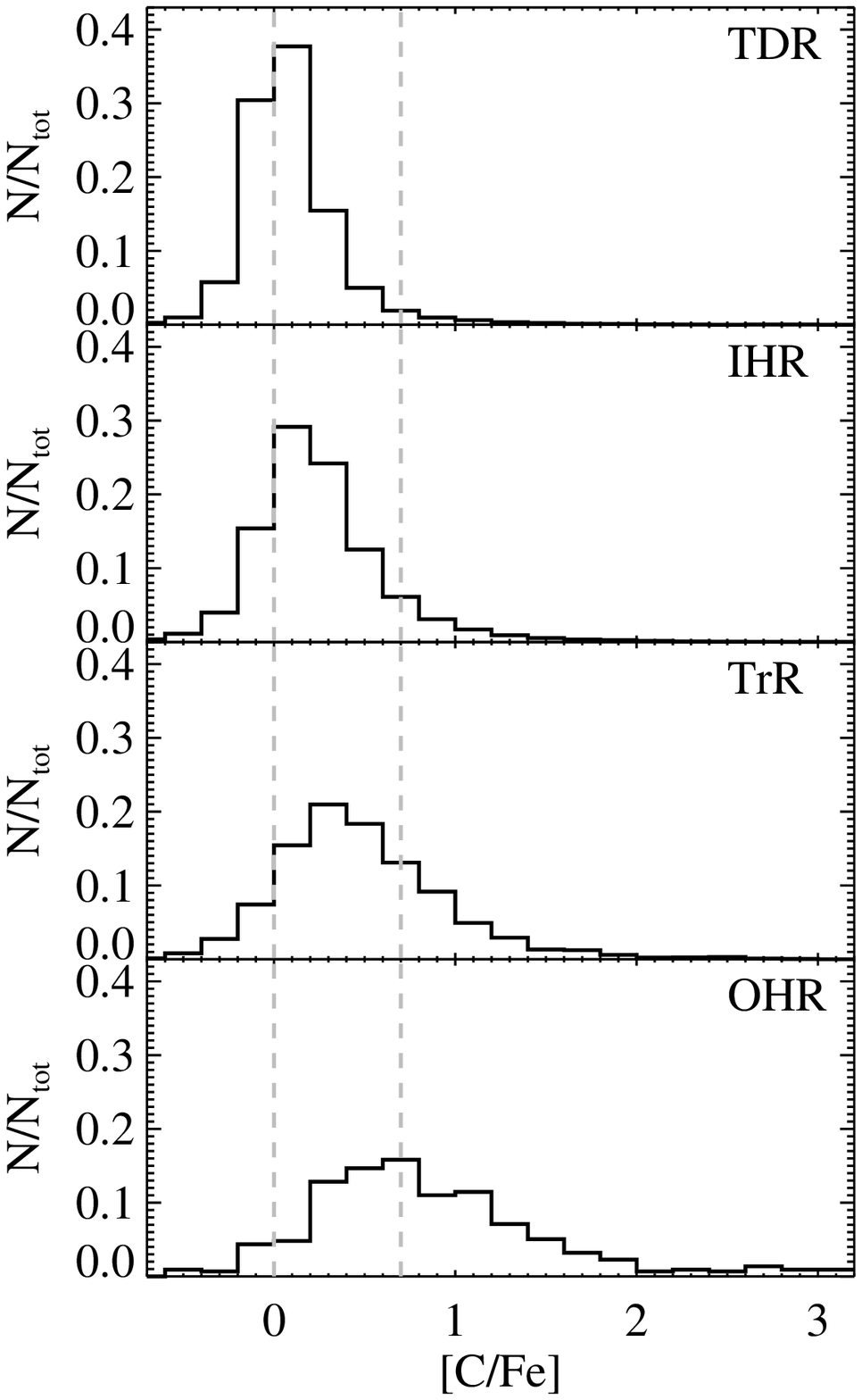}
\caption{Histograms of the normalized fractions of metallicities ([Fe/H], 
left panels) and carbonicities (\cfe, right panels) in the four regions we have
identified (see text). The labels TDR, IHR, TrR, and OHR represent the
thick-disk region, the inner-halo region, the transition region, and the
outer-halo region, respectively, defined on the basis of Figure
\ref{fig:cmap}. The total number ($N_{\rm tot}$)
of stars for each region is listed in each panel. The vertical
dashed lines in the left panels represent the mean metallicity of
the canonical thick disk, [Fe/H] = --0.6, the IHP, [Fe/H] = --1.6, and the OHP, [Fe/H] =
--2.2, according to \citet{carollo2007, carollo2010}. Gaussians derived from
a simple mixture-model analysis are shown in blue for the IHP, red for
the OHP, and green for the canonical TDP. The two vertical lines in the
right panels denote the solar \cfe\ and \cfe\ = $+$0.7. The
total number of stars in each region shown in the right panels
is the same as in the left panels.}
\label{fig:mdf}
\end{figure*}

We adopt the distance cut at \z\ $\leq$ 3 kpc for the TDR by considering the
scale heights of the thick disk and the metal-weak thick disk determined by
\citet{carollo2010}. The boundaries for the IHR and OHR are determined
by inspection of the carbonicity contours. We emphasize that our scheme
to separate the stellar populations for the IHR and OHR (as well as the
TrR) is $not$ based on metallicity or kinematics, but relies solely on
the level of [C/Fe] at a given location. These regions are shown
with white labels in Figure~\ref{fig:cmap}.

\subsection{Spatial Distribution of Metallicity, [Fe/H]}

Figure \ref{fig:fmap} is a metallicity map of our program sample over
the same range of \z\ and $R$ as shown in Figure \ref{fig:cmap}. The
regions identified in the carbonicity map are also labeled in this
figure. From inspection, each of the carbonicity regions exhibits a different 
level of metallicity -- the IHR with [Fe/H] $\sim$ --1.5, and
the OHR with [Fe/H] $\lesssim$ --2.0, similar to the proposed divisions
of the IHP and OHP obtained in the analysis of \citet{carollo2007,
carollo2010} on the basis of local kinematics (stars within 4 kpc of the
Sun). Note that the Monoceros Stream \citep{newberg2002, ivezic2008} is
clearly identifiable in the figure, and appears as the yellow-orange
region with [Fe/H] $\sim$ --1.0 at $ 10 < R < 20$ kpc and \z\ $< 5$ kpc.

\subsection{Dichotomy from the Metallicity Distribution Functions}

We now investigate the metallicity distributions of each region defined in Figure
\ref{fig:cmap}, to test whether we can identify distinct MDFs
associated with the inner- and outer-halo populations.

Figure \ref{fig:mdf} shows the MDFs (left column of panels) and
CDFs (right column of panels) for the TDR, IHR, TrR, and OHR. Note 
that, when selecting the stars in each
region, in addition to the criteria described in Figure~\ref{fig:cmap},
we applied \z\ $>$ 3 kpc for the IHR, \z\ $>$ 5 kpc for the TrR, and \z\
$>$ 7 kpc for the OHR, in order to minimize cross-contamination from
stars in the different regions.

Inspection of the left column of panels in the figure clearly shows that
the peak of the MDFs shifts to lower metallicity as we move from the IHR
to the OHR (from the second to the fourth plot). The MDF of the TDR
shows two peaks -- one arising from the canonical thick-disk
population at [Fe/H] $\sim$ --0.6, the other from the IHP at [Fe/H]
$\sim$ --1.5. The MDF for stars in the IHR exhibits a well-defined
distribution with a peak at [Fe/H] = --1.5, associated with the IHP. The
MDF of the TrR appears to have a broad distribution, presumably due to the
presence of overlapping contributions from the IHP and the OHP. Finally,
the stars in the OHR exhibit a well-defined peak at [Fe/H] $\sim$ --2.2.
associated with the OHP.

Our peak value of [Fe/H] = --1.5 for stars in the IHR, which we argue
are likely associated with the IHP, is similar to that obtained by \citet{an2015} for
the IHP, [Fe/H] = --1.4, while the peak we obtain for the
OHR, [Fe/H] = --2.2, is somewhat lower than the [Fe/H] = --1.9 for
the OHP found by \citet{an2015}. Note that An et al. explored an {\it in
situ} sample of main-sequence stars covering a similar distance range to 
our MSTO sample, but truncated beyond 10 kpc from the Sun. The
difference in the metallicity peaks of the OHP between the two analyses
likely arises due to the increasing lack of sensitivity of photometric metallicity
techniques based on broadband $ugriz$ for stars with [Fe/H] $< -2.0$.
Even though the divisions of the regions we consider were made on the
basis of carbonicity, the peaks of the MDFs for the dominant populations
we associate with each region also agree very well with those inferred
to exist from an analysis of local kinematics by \citet{carollo2007, carollo2010}.
Reaching such similar conclusions from three very
different approaches provides confidence in the reality of the dual-halo
nature of the MW.

The panels in the left column in Figure \ref{fig:mdf} also demonstrate the
clear advantage gained by using chemical-abundance ratios to distinguish one
component from another. In these panels, the green, blue, and red curves
represent individual MDFs evaluated from a simple two-component Gaussian
mixture-model analysis. The blue distribution represents the IHP, while
the red distribution is associated with the OHP.
Note that the stars assigned to the IHP by this approach have almost no
contribution from stars of the OHP, whereas the objects classified as the OHP
have only a negligible contribution from the IHP. These results indicate that
chemical-abundance ratios provide a more powerful discriminator for the
identification of stellar populations than the kinematic and orbital
information (including angular momentum and integrals of motion) alone, which
can result in substantial overlap of the populations.

The right column of panels in Figure \ref{fig:mdf} shows the CDF for
each region. The number of stars in each panel is the same as in the
corresponding panel on the left. Inspection of these panels indicates
that the fraction of stars with carbon enhancements relative to iron
clearly varies from region to region. The fiducial dashed lines indicate
the solar [C/Fe] and the level of carbon enhancement conventionally used
to identify CEMP stars, [C/Fe] $\ge +0.7$. The fraction of CEMP stars
increases as one proceeds from the TDR to the IHR, TrR, and OHR. This
provides suggestive, but not yet definitive, evidence that different
progenitors may be responsible for the majority production of carbon in
the underlying stellar populations we associate with these regions.

\begin{figure}
\centering
\plotone{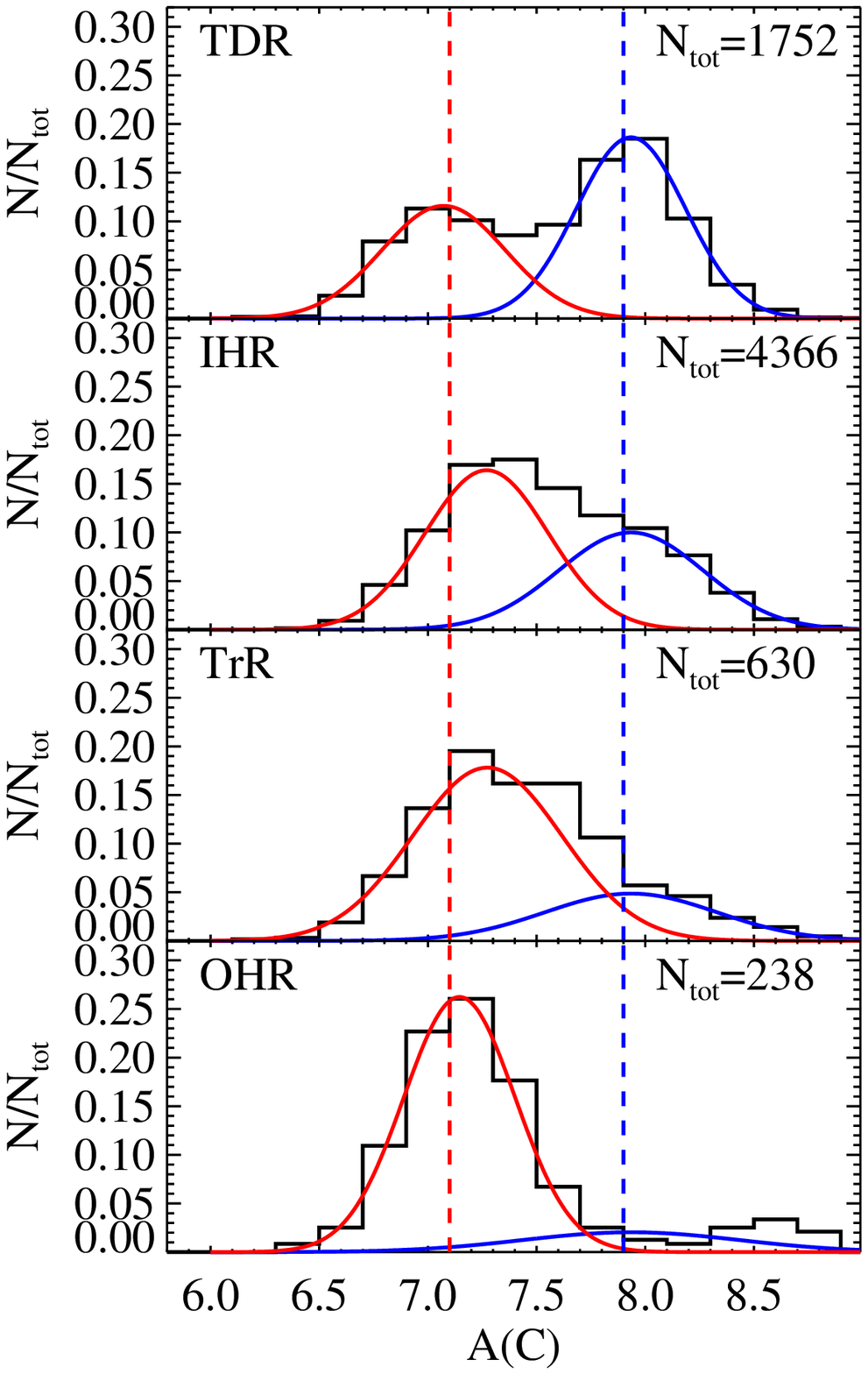}
\caption{Histograms of the normalized fractions of absolute carbon abundances,
$A$(C), for stars with [Fe/H] $\le -1.0$ and \cfe\ $\geq +0.7$. The blue
dashed vertical line represents the mean $A$(C) of the stars that \citet{yoon2016}
classify as CEMP-$s$, while the red dashed vertical line is the reference line to
distinguish between the CEMP-no and CEMP-$s$ stars used in \citet{yoon2016}. The
notation for TDR, IHR, TrR, and OHR is the same as in Figure
\ref{fig:mdf}. Gaussians for a simple mixture-model analysis of the CEMP stars are
shown in blue for the high-$A$(C) stars and red for the low-$A$(C) stars. The
total number ($N_{\rm tot}$) of stars used to make the histogram
is listed in each plot.}
\label{fig:ac}
\end{figure}

\subsection{Dichotomy from Distribution of Absolute Carbon Abundances, $A$(C)}

The diversity of the heavy-element abundance patterns
found among CEMP stars (CEMP-no, CEMP-$s$, CEMP-$r$, and CEMP-$r/s$)
already indicates that there is likely to be more than one source
of carbon production at low metallicity, and perhaps even within the
subclasses themselves (see the discussion by \citealt{yoon2016}). This,
along with the contrast seen in the carbonicity map of the
halo-system stars, suggests that different subclasses of CEMP stars
might be associated with different dominant nucleosynthesis pathways
within their associated stellar populations.

\citet{carollo2014} tested this idea using the limited number of
CEMP stars with high-resolution measurements of [C/Fe] and [Ba/Fe] then
available to calculate the relative fractions of CEMP-$s$ and CEMP-no
stars associated with the IHP and OHP, separated on the basis of their
integrals of motion. These authors demonstrated that the CEMP stars
associated with the OHP exhibit a higher fraction (by about a factor of two)
of CEMP-no stars than CEMP-$s$ stars, while the IHP only
slightly favors CEMP-$s$ stars over CEMP-no stars. A much larger sample
of CEMP stars is required, however, for confirmation of this
behavior.

According to the conventional criterion used to distinguish CEMP-$s$
stars from CEMP-no stars, a high-resolution spectroscopic measurement of
(at least) the [Ba/Fe] ratio is required.  However, \citet{yoon2016}
have recently developed a new method for separation of these subclasses
of CEMP stars based solely on the absolute carbon abundance, $A$(C),
which can be obtained from medium-resolution spectroscopy such as that
available for our SDSS program stars. According to the classification
exercise carried out by \citet{yoon2016}, CEMP-$s$ stars have
\ac\ $>$ 7.1, while the CEMP-no stars have \ac\ $\le$ 7.1.
We consider this approach to test whether or not the
relative fractions of CEMP-no and CEMP-$s$ stars in the OHR differ
significantly from those in the IHR, and by inference, between the OHP
and the IHP.

Figure~\ref{fig:ac} shows the distribution of \ac\ for each region of
the carbonicity map in Figure \ref{fig:cmap}. To compare the two
different subclasses of CEMP stars, we include only the stars with
\cfe\ $\geq +0.7$ and \feh\ $\leq$ --1.0 in the figure. The blue and red curves
are Gaussians obtained from a simple two-sample mixture model for each
region. The blue dashed vertical line represents the mean $A$(C) of the
stars that Yoon et al. classify as CEMP-$s$, while the red dashed vertical
line is the reference line to distinguish between the CEMP-no and
CEMP-$s$ stars used by \citet{yoon2016}. From inspection of this figure,
the IHR exhibits a very broad distribution with likely separable peaks
at \ac\ $\sim$ 7.3 and $A$(C) $\sim$ 7.9, respectively, while the OHR
exhibits a strong peak at \ac\ = 7.1 and only a very small peak at
$A$(C) $\sim$ 7.9. Inspection of the $A$(C) distribution in the TrR
shows the expected shift between these two different behaviors, because the
relative proportion of the low-\ac\ stars increases as one moves from the
IHR to the OHR.

Quantitatively, if we separate stars in each region at \ac\ = 7.1, the
dividing line between CEMP-no and CEMP-$s$ according to
\citet{yoon2016}, we obtain fractions of 84\% and 16\% for high-\ac\
and low-\ac\ stars in the IHR, 77\% and 23\% in the TrR, and 63\% and
37\% in the OHR, respectively. On the other hand, using the fractions of
high- and low-\ac\ stars identified by the Gaussian mixture model for
each region, we obtain 42\% for the high-\ac\ stars (blue curve in the
figure) and 58\% for the low-\ac\ stars (red curve in the figure) in the
IHR, 24\% and 76\% in the TrR, and 13\% and 87\% in the OHP,
respectively. These results clearly indicate an increasing trend in the
relative fractions of the low-\ac\ stars with respect to the high-\ac\
stars from the IHR to the OHR. The increasing fraction 
of low-\ac\ stars provides strong suggestive evidence that 
different progenitors contributed to stellar populations we 
associate with these regions.

Unlike the mean value of \ac\ = 6.3 for the low-\ac\ stars found by
\citet{yoon2016}, represented by CEMP-no stars (classifications
based on high-resolution spectroscopy for a sample that included
numerous giants), our distribution of low-\ac\ stars exhibits a peak
around \ac\ = 7.1. This is $not$ primarily because of our inability to
detect low-\ac\ CEMP-no stars in our sample of metal-poor MSTO stars,
$but~rather$ because the stars in our program sample cover
substantially different ranges of metallicity and luminosity. As shown
in Figure 1 of \citet{yoon2016}, which we refer to as the Yoon--Beers
diagram, their CEMP-no stars are dominated by objects with [Fe/H] $<$
--3.0. On the other hand, as seen from Figure~\ref{fig:mdf}, our sample
of MSTO stars is dominated by objects with [Fe/H] $>$ --2.5. The
Yoon--Beers diagram of \citet{yoon2016} reveals that most of the CEMP-no
stars in this metallicity range have \ac\ $>$ 6.3. If we consider only 
the CEMP-no stars with [Fe/H] $>$ --3.0 from their sample, we obtain a
peak for the low-\ac\ stars at \ac\ $\sim$ 6.6. Moreover, excluding the
giants with \logg\ $< 3.5$ in the sample of \citet{yoon2016}, which are
not present in our sample, we obtain a low-\ac\ peak value of \ac\ = 6.8.
These two values are much closer to our determinations. The high-\ac\
stars from Yoon et al., which include more objects with [Fe/H] $>$
--2.5, have a mean value of \ac\ = 7.9, in excellent agreement with our
peak value of \ac\ $\sim$ 7.9, shown by the blue curve in
Figure~\ref{fig:ac}.

This exercise raises two important points. First, the appropriate $A$(C)
division line between CEMP-$s$ and CEMP-no stars may depend, at least
weakly, on stellar temperature and luminosity class, and secondly, the
the peak value of \ac\ for CEMP-no stars varies over the metallicity
range considered in a given sample. A value of $A$(C) $\sim 7.6$ would be
a more suitable location to divide the low- and high- \ac\ stars for our
MSTO sample, rather than the level of $A$(C) = 7.1 for the sample of 
the Yoon et al., which included substantial numbers of cooler stars and 
more metal-poor stars.

\section{Implications for the Origin of the Duality in the Galactic Halo System}

As discussed above, when the halo populations of the MW are separated
based on their carbonicity, [C/Fe], we find that the component
associated with the IHP exhibits a peak metallicity at [Fe/H] $\sim$
--1.5, while the peak metallicity associated with the OHP is [Fe/H]
$\sim$ --2.2. These results are completely consistent with previous
studies by \citet{carollo2007, carollo2010} and \citet{an2013, an2015}.
We have also quantitatively assessed the fraction of CEMP-no and
CEMP-$s$ stars in the OHP and IHP, classified on the basis of their
observed \ac, and found that the stars associated with the OHP exhibit a
higher proportion of CEMP-no stars, compared to CEMP-$s$ stars, than
found for stars associated with the IHP. These results suggest that the
progenitors of these two classes of CEMP stars contributed to the outer-
and inner-halo populations in different proportions, a valuable clue to
unraveling the apparently different assembly histories of the OHP and
IHP.

Current simulations of the hierarchical assembly of MW-like galaxies
within the $\Lambda$-CDM paradigm (e.g., \citealt{tissera2012,
tissera2013, tissera2014}) show that the inner halo formed through
mergers of a few relatively massive gas-rich mini-halos, which underwent
dissipative (angular-momentum-conserving) collapse, in contrast with the
stars of the outer halo, which formed in numerous lower-mass mini-halos
that were accreted into the Galactic halo through chaotic,
dissipationless mergers. In such a scenario, star formation in the
relatively massive gas-rich subfragments that led to the inner halo
would be expected to be very efficient, and chemical enrichment could
proceed rapidly. This would lead to the formation of moderately
metal-poor stars with [Fe/H] $\sim$ --1.5, including numerous
intermediate-mass ($\sim$ 1--4 \msun) asymptotic giant branch (AGB)
stars, a subset of which formed as members of the binaries that produced
(via mass transfer of their nucleosynthesis products) the CEMP-$s$ stars
observed today. CEMP-no stars would have formed from the nucleosynthesis
products of more massive progenitors during the first, or at least
early, bursts of star formation in their parent mini-halos; some might
also be expected to be present among early-generation stars in the
mini-halos associated with the IHP. In this manner, one might understand
the preferential production of CEMP-no stars at low metallicity,
relative to CEMP-$s$ stars, as reported by previous studies (e.g.,
\citealt{aoki2007, placco2014, yoon2016}), as well as our present work.

The CEMP-no and CEMP-$s$ stars are thought to originate
from different classes of astrophysical progenitors: (1) CEMP-no stars 
from the so-called ``faint supernovae'' or ``mixing-and-fallback'' models
with masses of $\sim$20--60 \msun\ (e.g., \citealt{umeda2003,
umeda2005, nomoto2013, tominaga2014}), or ``spinstar'' models with masses
$>$ 60--100 \msun\ (\citealt{meynet2006, meynet2010, chiappini2013}); 
(2) CEMP-$s$ stars from binary mass transfer from intermediate-mass
(1--4 \msun) AGB stars, which have now evolved to become faint white
dwarfs (e.g., \citealt{suda2004, herwig2005, lucatello2005, komiya2007,
bisterzo2011, hansen2015}). The results from recently completed
long-term radial-velocity monitoring programs (\citealt{starkenburg2014,
hansen2016a, hansen2016b, jorissen2016}) provide strong support for
these assertions, based on the very different binary fractions of 
these two subclasses of CEMP stars.

Since the proposed mass ranges for the progenitors of CEMP-no and
\cemps\ differ substantially, the relative fractions of CEMP-no and \cemps\
stars between the IHP and OHP can potentially provide strong constraints on the
initial mass functions (IMFs) in the environments that were dominantly
responsible for the formation of the two populations, as suggested by a
number of previous authors (e.g., \citealt{carollo2014, lee2014}, and
references therein). The best surviving candidates for the environments
in which early star formation took place (perhaps corresponding to the
lower-mass mini-halos that contributed stars to the OHP) are the
ultra-faint dwarf (UFD) galaxies around the MW discovered during the
SDSS (e.g., \citealt{belokurov2006a, belokurov2006b, zucker2006}). It is
presumably no coincidence that our substantially higher fraction of
CEMP-no stars associated with the OHP, compared with the CEMP-$s$ stars
of the OHP, is consistent with the relative dominance of CEMP-no stars
among the UFDs associated with the MW (e.g., \citealt{frebel2014}). Indeed,
the large fraction of CEMP-no stars we
associate with the OHP suggests that systems similar to disrupted UFD
galaxies may have contributed significantly to the formation of the
OHP of the MW.

\section{Summary and Conclusions}

We have used a sample of over 100,000 MSTO stars from the SDSS to
investigate the $in~situ$ spatial distribution of carbonicity ([C/Fe])
of the Galactic halo system, and demonstrated that it exhibits different
levels of carbonicity as the distance from the Galactic plane increases.
Dividing our sample into four regions in the plane of \z\ vs. $R$, based on
their carbon enhancement and location, we construct the MDF for each
region. The MDFs of the halo regions are interpreted in terms of the
differing contributions from the inner- and outer-halo populations. The
IHP exhibits a peak at [Fe/H] $\sim$ --1.5, while a peak at [Fe/H]
$\sim$ --2.2 is associated with the OHP, consistent with a number of
previous studies. This result provides further evidence for the duality
of the Galactic halo, based on an entirely different approach from the
ones used before.

We have quantitatively assessed the relative fractions of CEMP-no and
CEMP-$s$ stars (classified on the basis of their derived absolute carbon
abundances, $A$(C)) in the IHR and OHR (which we associate with dominant
contributions from the IHP and OHP), and find that that the CEMP stars
in the OHR exhibit a significantly higher proportion of CEMP-no stars
relative to CEMP-$s$ stars than found for the CEMP stars in the IHR.

The different MDFs and CEMP-no/CEMP-$s$ fractions between the stars we
associate with the IHP and OHP suggest that these individual halo components
have received contributions of the nucleosynthesis products from
different astrophysical progenitors, and experienced different assembly
histories. Ultimately, this information can be used to constrain the
mass distributions of the mini-halos from which they formed, as well as
their star-formation histories.

We are currently exploring maps of the spatial distributions for larger
samples of CEMP-no and CEMP-$s$ stars, based on their identification
from the Yoon--Beers diagram of $A$(C) vs. [Fe/H] (Figure 1 of
\citealt{yoon2016}), drawing on both the SDSS and the Large Sky Area
Multi-Object Fiber Spectroscopic Telescope (LAMOST;
\citealt{cui2012}) surveys.  Such maps can be compared with current and
future numerical simulations of the formation of the halo of the MW. A
new analysis of the kinematics for much larger samples of CEMP-no and
CEMP-$s$ stars than previously considered, based on more recent high-
and medium-resolution spectroscopic analyses, is also now underway, and
should prove illuminating.

\acknowledgments

We thank an anonymous referee for his/her careful review of this
manuscript, and for pointing out a number of places where we could
improve the clarity of the presentation.

Funding for SDSS-III has been provided by the Alfred P. Sloan Foundation, the
Participating Institutions, the National Science Foundation, and the U.S.
Department of Energy Office of Science. The SDSS-III Web site is
http://www.sdss3.org/.

SDSS-III is managed by the Astrophysical Research Consortium for the
Participating Institutions of the SDSS-III Collaboration including
the University of Arizona, the Brazilian Participation Group,
Brookhaven National Laboratory, University of Cambridge, Carnegie
Mellon University, University of Florida, the French Participation
Group, the German Participation Group, Harvard University, the
Instituto de Astrofisica de Canarias, the Michigan State/Notre
Dame/JINA Participation Group, Johns Hopkins University, Lawrence
Berkeley National Laboratory, Max Planck Institute for Astrophysics,
Max Planck Institute for Extraterrestrial Physics, New Mexico State
University, New York University, Ohio State University, Pennsylvania
State University, University of Portsmouth, Princeton University,
the Spanish Participation Group, University of Tokyo, University of
Utah, Vanderbilt University, University of Virginia, University of
Washington, and Yale University.

This work was supported by the 2014 research fund of Chungnam National
University. Y.S.L. also acknowledges partial support from the National
Research Foundation of Korea to the Center for Galaxy Evolution Research
and Basic Science Research Program through the National Research
Foundation of Korea (NRF) funded by the Ministry of Science, ICT and 
Future Planning (NRF-2015R1C1A1A02036658). T.C.B., V.P., J.Y., and D.C.
acknowledge partial support for this work from grants PHY 08-22648:
Physics Frontier Center/Joint Institute of Nuclear Astrophysics (JINA),
and PHY 14-30152: Physics Frontier Center/JINA Center for the Evolution
of the Elements (JINA-CEE), awarded by the US National Science
Foundation. T.M. acknowledges support by the European Union FP7 programme
through ERC grant number 320360. This work benefited from support by the
US National Science Foundation under Grant No. PHY 14-30152 (JINA Center
for the Evolution of the Elements).

\clearpage

\end{document}